\documentclass{article} 
\usepackage[a4paper]{geometry} % Adjust margins
\usepackage{setspace}
\usepackage{authblk}
\usepackage{natbib}
\usepackage{graphicx}
\usepackage{xcolor}
\usepackage{float}
\usepackage{caption}
\usepackage{pdfpages}

\newcommand{\IB}{\ensuremath{\mathbf{I}}}

\newcommand{\BBs}{\ensuremath{\mathbf{B}_{\mathrm{sp}}}}
\newcommand{\SB}{\ensuremath{\mathbf{S}}}

\newcommand{\MB}{\ensuremath{\mathbf{M}}}
\newcommand{\UB}{\ensuremath{\mathbf{U}}}

\newcommand{\xb}{\ensuremath{\mathbf{x}}}
\newcommand{\yb}{\ensuremath{\mathbf{y}}}
\newcommand{\zb}{\ensuremath{\mathbf{z}}}
\newcommand{\ub}{\ensuremath{\mathbf{u}}}
\newcommand{\vb}{\ensuremath{\mathbf{v}}}

\newcommand{\rb}{\ensuremath{\mathbf{r}}}
\newcommand{\ssb}{\ensuremath{\mathbf{s}}}
\newcommand{\zerob}{\ensuremath{\mathbf{0}}}

\newcommand{\betah}{\ensuremath{\widehat{\beta}}}

\newcommand{\bbetas}{\ensuremath{\boldsymbol{\beta}_{\mathrm{sp}}}}
\newcommand{\eepsilon}{\ensuremath{\boldsymbol{\epsilon}}}
\newcommand{\xxi}{\ensuremath{\boldsymbol{\xi}}}
\newcommand{\sigmainv}{\ensuremath{\sigma^{-2}}}
\newcommand{\RR}{\ensuremath{\mathbb{R}}}
\newcommand{\BSigma}{\ensuremath{\boldsymbol{\mSigma}}}

\newcommand{\gammab}{\ensuremath{\boldsymbol{\gamma}}}

\newcommand{\BBstilde}{\ensuremath{\tilde{\mathbf{B}}_{\mathrm{sp}}}}
\newcommand{\bbetastilde}{\ensuremath{\tilde{\boldsymbol{\beta}}_{\mathrm{sp}}}}

\usepackage{amsmath}

\usepackage{times}
\usepackage{bm}
\usepackage{natbib}
\usepackage{amsfonts}
\usepackage{paralist}
\usepackage{xr}
\usepackage{xcolor}

 \DeclareMathOperator{\E}{E}

\def \mSigma   {\mathbf{\Sigma}}

\newtheorem{proposition}{Proposition}
\newtheorem{lemma}{Lemma}

\newtheorem{corollary}{Corollary}

\makeatletter
\newcommand*{\addFileDependency}[1]{% argument=file name and extension
\typeout{(#1)}% latexmk will find this if $recorder=0

\@addtofilelist{#1}

\IfFileExists{#1}{}{\typeout{No file #1.}}
}\makeatother

\title{Demystifying Spatial Confounding}

\author{
Emiko Dupont$^{1}$(eahd20@bath.ac.uk),  Isa Marques$^{2}$, 
and Thomas Kneib$^{3}$, \\
$^{1}$Department of Mathematical Sciences, University of Bath, \\ Claverton Down,  Bath BA2 7AY, U.K.\\
$^{2}$ Department of Statistics, The Ohio State University, \\ Cockins Hall, 1958 Neil Ave, Columbus 43210, U.S.A. \\
$^{3}$Chair of Statistics and Campus Institute Data Science, University of G{\"o}ttingen, \\  Humboldtallee 3, 37073 G{\"o}ttingen, Germany
}

\date{}

\begin{document}

\maketitle

\begin{abstract}
%Has to be a single paragraph, at most 225 words.
Spatial confounding is a fundamental issue in spatial regression models which arises because spatial random effects, included to approximate unmeasured spatial variation, are typically not independent of covariates in the model. This can lead to significant bias in covariate effect estimates. The problem is complex and has been the topic of extensive research with sometimes puzzling and seemingly contradictory results. Here, we develop a broad theoretical framework that brings mathematical clarity to the mechanisms of spatial confounding, deriving an explicit analytical expression for the resulting bias. We see that the problem is directly linked to spatial smoothing and identify exactly how the size and occurrence of bias relate to the features of the spatial model as well as the underlying confounding scenario. Using our results, we can explain subtle and counter-intuitive behaviours. Finally, we propose a general approach for dealing with spatial confounding bias in practice, applicable for any spatial model specification. When a covariate has non-spatial information, we show that a general form of the so-called spatial+ method can be used to eliminate bias. When no such information is present, the situation is more challenging but, under the assumption of unconfounded high frequencies, we develop a procedure in which multiple capped versions of spatial+ are applied to assess the bias in this case. We illustrate our approach with an application to air temperature in Germany.
\end{abstract}

% %  Please place your key words in alphabetical order, separated
% %  by semicolons, with the first letter of the first word capitalized,
% %  and a period at the end of the list.
% %

\begin{center}
\end{center}

% %  As usual, the \maketitle command creates the title and author/affiliations
% %  display 

\maketitle

% %  If you are using the referee option, a new page, numbered page 1, will
% %  start after the summary and keywords.  The page numbers thus count the
% %  number of pages of your manuscript in the preferred submission style.
% %  Remember, ``Normally, regular papers exceeding 25 pages and Reader Reaction 
% %  papers exceeding 12 pages in (the preferred style) will be returned to 
% %  the authors without review. The page limit includes acknowledgements, 
% %  references, and appendices, but not tables and figures. The page count does 
% %  not include the title page and abstract. A maximum of six (6) tables or 
% %  figures combined is often required.''

% %  You may now place the substance of your manuscript here.  Please use
% %  the \section, \subsection, etc commands as described in the user guide.
% %  Please use \label and \ref commands to cross-reference sections, equations,
% %  tables, figures, etc.
% %
% %  Please DO NOT attempt to reformat the style of equation numbering!
% %  For that matter, please do not attempt to redefine anything!

\section{Introduction}
\label{s:intro}
%Spatial statistics is concerned with the analysis of spatially indexed data, usually using regression models of the form 
In spatial statistics, regression models usually take the form
\begin{equation}
 y(s) = \beta x(s) + \gamma(s) + \epsilon(s)
\label{eq:spatreg}    
\end{equation}
where $y(s)$ denotes the response variable of interest, $x(s)$ a covariate with regression coefficient $\beta$, $\gamma(s)$ a spatially correlated random effect and $\epsilon(s)$ independent and identically distributed error terms, all evaluated at locations $s$ within some spatial domain. %We consider the case of only one covariate but the model could be expanded to include more. 
The spatial effect $\gamma(s)$ is included %to avoid model mis-specification by approximating unmeasured spatial variation in the response variable which typically arises due to unobserved or unknown spatial variables.
to approximate unmeasured spatial variation in the response which typically arises due to unobserved spatial variables.
Specific instances of the model are obtained by specifying the type of spatial information (e.g.\ discrete regional vs.\ continuous coordinate-based) and the representation of the spatial effect $\gamma(s)$ (e.g.\ Gaussian random fields, Markov random fields, thin plate splines, etc.).

Although not always explicitly stated, the regression \eqref{eq:spatreg} assumes that the covariate $x(s)$ is given and fixed. This implies independence between the covariate and the spatial effect, which is unlikely to hold for observational data where covariates are realisations of potentially spatially dependent random variables. If indeed this assumption is violated, it leads to a phenomenon known as spatial confounding, where the estimate of the covariate effect $\beta$ becomes biased even when the model is correctly specified.
%The regression \eqref{eq:spatreg} almost always assumes independence between the spatial effect $\gamma(s)$ and the error terms $\epsilon(s)$ and, although not always explicitly stated, %relies on the typical modelling assumption that the covariate that $x(s)$ is given and fixed. 
%\textcolor{magenta}{that the covariate $x(s)$ is given and fixed.} The latter, however, implies independence between the covariate and the spatial effect, which is unlikely to hold for observational data where covariates are realisations of potentially spatially dependent random variables. If indeed this assumption is violated, it leads to a phenomenon known as spatial confounding where the estimate of the covariate effect $\beta$ becomes biased even when the model is correctly specified.
%Violating this assumption leads to a phenomenon known as spatial confounding where the estimate of the covariate effect $\beta$ becomes biased even when the model is correctly specified.
Thus, understanding spatial confounding is fundamental to reliable estimation of covariate effects in spatial statistics, and the problem has received considerable attention in the literature since the seminal work of \citet{clayton1993spatial} and, later, \citet{hodges2010adding}. 
Recently, there has been a surge in interest, not least because the established understanding and methods most commonly used for dealing with the problem have been shown to be problematic. 

Broadly, research into spatial confounding can be structured along the following lines: (i) Investigations and suggestions for potential solutions under specific instances of the model \eqref{eq:spatreg} such as a choice of discrete or continuous spatial dependence and/or a particular representation of the spatial effect \citep{dupont2021spatial+, guan2020spectral, marques2022mitigating,urdangarin2022evaluating}. (ii) Identifying factors that influence the size or occurrence of bias, in particular, the importance of the spatial scales of both measured and unmeasured variables % such as spatial scales or smoothness of both measured and unmeasured variables, or a mismatch between the model specification and the true data generating process 
\citep{paciorek2010importance,page2017estimation,keller2020selecting, nobre2020effects, bobb2022accounting}. (iii) Research based on a (sometimes implicit) preference for the estimate of $\beta$ in a non-spatial analysis \citep{reich2006effects, hanks2015restricted, hughes2013dimension, hefley2017bayesian, adin2021alleviating, briz2023alleviating}. This approach, usually involving a method known as restricted spatial regression, dominated the literature as well as practical implementations for many years, but has since been shown to be problematic for several reasons \citep{khan2020restricted, zimmerman2021deconfounding, bradley2024spatial}. %Indeed, while the ordinary least squares estimate in a correctly specified non-spatial analysis is unbiased, if the data generating process includes unmeasured spatial confounders, the assumptions of the non-spatial model are violated and, therefore, unbiasedness no longer holds. 
(iv) Research based on causal inference approaches \citep{thaden2018structural, papadogeorgou2019adjusting, schnell2020mitigating}, an overview of which is in \citet{reich2020review}. %Our focus in this paper is on understanding the mechanisms of confounding from a mathematical point of view, which can be used in conjunction with causal inference considerations. 

%Although the many recent contributions to the literature have shed more light on spatial confounding, results appear at times to be counter-intuitive or mutually conflicting, making it difficult to draw meaningful conclusions. 
Although this work has shed more light on spatial confounding, results appear at times to be counter-intuitive or mutually conflicting. %, making it difficult to draw definitive conclusions.
This is in part because the complexity of spatial models means that conclusions are often based on simulation studies, potentially
%dependent on the choices made for the simulation setup.
sensitive to the specific simulation setup. Moreover, the research direction (iii) above has led to a focus on the difference between the estimates in the spatial and non-spatial analyses for identifying confounding bias. However, as our results confirm, this difference does not in itself determine whether either estimate is biased.
%In addition to the above, a number of recent works (Gilbert, Bolin, Dupont) have investigated the asymptotic behaviour of covariate effect estimates in models of the form (\ref{eq:spatreg}), in particular, Gilbert and Dupont show that the bias tends to zero as the sample size goes to infinity. In this paper we focus on the behaviour of the covariate effect estimate in (\ref{eq:spatreg}) under a finite sample size. Using our results, we show that there are scenarios in which spatial confounding bias is negligible (in line with \citep{khan2023re}), however, there are also scenarios in which the bias can become arbitrarily large. }

%this is not directly related to the occurrence or otherwise of bias. There have also been attempts to prove that spatial confounding bias in models of the form (\ref{eq:spatreg}) is always negligible so that the issue does not exist in practice \citep{khan2023re}, but we will see that bias from spatial confounding can in fact become arbitrarily large. %While this may be an appealing result, we will see that bias from spatial confounding can in fact become arbitrarily large. }

%\subsection*{Our contribution}

We develop a theoretical framework that brings mathematical clarity to the mechanisms of spatial confounding and the methods used for dealing with the resulting bias. We show that it is a problem for spatial models in general, directly linked to spatial smoothing \citep[in line with work on bias  
in linear mixed models,][]{schnell2019spectral}. %but derived and interpreted specifically for the spatial context. %\textcolor{magenta}{Our results are in line with work on bias %in fixed effect estimates in linear mixed models \citep{schnell2019spectral} but derived and interpreted specifically for the spatial context.
Central to our findings is the analytical expression for the bias derived in Proposition \ref{prop:bias} which shows that, in the metric defined by the precision matrix of the chosen spatial analysis model, the bias is {\em the correlation between the covariate and the confounder times the size of the confounder relative to the size of the covariate}. Not only is this an intuitive result, %that aligns with similar results in other forms of regularized regression, 
but by using an eigendecomposition of the precision matrix, we derive a nuanced and detailed understanding of how the bias depends on the underlying confounding scenario and the covariance structure and parameters of the analysis model. This includes a deepened understanding of the role of spatial frequencies, defined directly in terms of the analysis model and the resulting eigendecomposition. %, and our conclusions are in line with \citet{paciorek2010importance,page2017estimation}. 
While the expression for the bias is itself relatively simple to derive and aligns with earlier results on regularized regression, the dependence on individual components and parameters in the spatial context is complex, explaining the subtle and at times counter-intuitive behaviours. We illustrate our findings in a simulation study of several different scenarios. 

A number of recent works \citep{dupont2021spatial+, gilbert2024consistency, bolin2024spatial} %(Gilbert, Bolin, Dupont) 
have investigated the asymptotic behaviour of the covariate effect estimate in models of the form (\ref{eq:spatreg}) under infill asymptotics, in particular, \citet{gilbert2024consistency} and \citet{dupont2021spatial+} show that if the covariate has non-spatial information, the bias tends to zero as the sample size goes to infinity. Of course, in practice, this limit is never reached. Our focus in this paper is on the behaviour of the covariate effect estimate in (\ref{eq:spatreg}) for a finite sample size. We show that, in this case, there are scenarios in which spatial confounding bias is negligible (in line with \citet{khan2023re}), however, there are also scenarios in which the bias can become arbitrarily large.

Based on our results, we develop a general framework for assessing and adjusting spatial confounding bias in practice, applicable under any specification of the spatial effects in \eqref{eq:spatreg}. The proposed method depends on whether or not the covariate has non-spatial information. %, as this determines whether the analysis model is identifiable. 
When non-spatial information is present, we show that a general form of spatial+ \citep{dupont2021spatial+} can be used to eliminate bias. When no such information is present, the situation is more challenging as further assumptions for identifiability are needed. Under the assumption of unconfounded high frequencies (as proposed in \citet{guan2020spectral}), we develop a method that utilises the spatial frequency behaviours of the covariate to assess the bias in this case. Specifically, we propose a procedure in which multiple capped versions of spatial+ are applied. The method can easily be adapted to the scenario in which other frequencies than the highest ones are assumed to be unconfounded.

We illustrate our approach with an application to monthly mean air temperature across Germany during the year 2010 with rainfall as covariate. Both variables display varying spatial frequency behaviours for different months of the year. We show how our methods can be applied to assess whether the effect estimate in the spatial model is biased and, if so, determine the estimated bias.

In the following, we first establish some preliminary results on spatial regression models in Section~\ref{sec:defs} that will allow us to derive compact and intuitive expressions for bias arising due to spatial confounding in Section~\ref{sec:bias}, supplemented by detailed simulations in Section~\ref{sec:simulations}. In Section~\ref{sec:adj}, we propose a method for assessing bias in practice, supplemented by an illustrative simulation study. Finally, in Section~\ref{sec:application}, we illustrate our approach with an application.

\section{Definitions and preliminary results}\label{sec:defs}

\subsection{Model specification}

Throughout this paper we assume that the true data generating process is given by
\begin{equation}\label{eqn:dgp}
 \yb = \beta\xb + \zb + \eepsilon, \quad \eepsilon\sim N(\zerob,\sigma^2\IB)
\end{equation}
where $\yb=(y(s_1),\ldots,y(s_n))^T$, $\xb=(x(s_1),\ldots,x(s_n))^T$ and $\eepsilon=(\epsilon(s_1),\ldots,\epsilon(s_n))^T$ are the response, covariate and error term at locations $s_1,\ldots,s_n$ and $\zb=(z(s_1),\ldots,z(s_n))^T$ the true confounding variable at these locations. Since the confounder is assumed to be unknown, in the spatial analysis model chosen by the researcher analysing the data, we assume that $\zb(s)$ is approximated by a spatial effect $\gamma(s)$ as in \eqref{eq:spatreg}. This implies an analysis model of the form

\begin{equation}\label{eqn:model}
\yb = \beta\xb + \BBs\bbetas+\eepsilon,\quad \bbetas\sim N(\zerob,\lambda^{-1}\SB^{-}),\text{ } \eepsilon\sim N(\zerob,\sigma^2\IB)
\end{equation}
where $\BBs$ is an $n\times p$ design matrix for the basis representing the spatial effect $\gamma(s)$, and $\bbetas$ the coefficients for the spatial effect in this basis. 
The matrix $\SB$ (with pseudo-inverse $\SB^{-}$) is a $p\times p$ penalty matrix which, together with the spatial basis in the design matrix $\BBs$, defines the structure of the spatial effect, and $\lambda>0$ is a smoothing parameter which controls the overall level of smoothing of the spatial effect. If the model matrix is identifiable (a discussion of this is in Section~\ref{sec:adj}), we can also allow the value $\lambda=0$, in which case the spatial random effect becomes unsmoothed additional fixed effect terms in the model. We assume that $\SB$ is symmetric positive semi-definite, i.e.\ it has non-negative eigenvalues $\alpha_1\le\cdots\le\alpha_p$. The null space of $\SB$, which could be non-trivial, corresponds to unsmoothed spatial vectors. The intercept in model (\ref{eqn:model}) is not included as a separate fixed effect, as it can be modelled as part of the spatial effect by including the vector $(1, \ldots, 1)^T$ as an unpenalised spatial basis vector in $\BBs$.

For comparison, we also consider the non-spatial analysis model
\begin{equation}\label{eqn:model_ns}
\yb = \beta_0+\beta\xb + \eepsilon,\quad \eepsilon\sim N(\zerob,\sigma^2\IB)
\end{equation}
where $\beta_0 \in \RR$ is an unknown intercept. 

Owing to the generality of our setup, all the commonly used spatial models fit into this generic framework. For example, for (i) Gaussian random fields based on coordinate information, $\BBs$ is an incidence matrix linking observations to observation locations and $\SB$ is the inverse covariance matrix implied by the covariance function (ii) (intrinsic) conditional auto-regressive and Gaussian Markov random field specifications for regional data, $\BBs$ is an indicator matrix for the regions and $\SB$ encodes spatial proximity with an adjacency matrix or graph Laplacian, and (iii) tensor product P-splines and thin plate splines, $\BBs$ contains basis function evaluations and $\SB$ the corresponding smoothness penalty. We can also take a Bayesian perspective if we have a multivariate Gaussian prior on $\bbetas$ with precision $\lambda\SB$, see \cite{fahkne11} for details on both the various options for spatial smoothing and the Bayesian perspective on spatial regression. 

\subsection{The spatial precision matrix $\BSigma^{-1}$}\label{sec:precision}
Central to our further analyses are the properties of the precision matrix $\BSigma^{-1}$ of the spatial analysis model \eqref{eqn:model}, derived from the covariance structure $\BSigma = \sigma^2\IB+\lambda^{-1}\BBs\SB^{-}\BBs^{T}$. Details of our derivations are included in Appendix~1.1. The following lemma shows how the eigenvectors of $\BSigma^{-1}$ are linked to the penalisation (i.e. the spatial smoothing) in the chosen analysis model.% Appendix~\ref{app:precision}.

\begin{lemma}\label{lem:Sigmainv_eig}
Let $\alpha_1\le\cdots\le\alpha_p$ be the eigenvalues of the penalty matrix $\SB$ and $\lambda>0$ the smoothing parameter. Then the eigenvalues of the precision matrix $\BSigma^{-1}$ are given by \linebreak $\{\sigmainv,\sigmainv w_1,\ldots,\sigmainv w_p\}$ where
$
w_i=\lambda\alpha_i/(\sigmainv+\lambda\alpha_i)\textrm{ for }i=1,\ldots, p.
$
\end{lemma}

Using this result, we obtain a natural decomposition of the sample space $\RR^n$. Specifically, Lemma~\ref{lem:Sigmainv_eig} shows that there exists an $n\times n$-matrix $\UB$ such that
\[
\UB^T\BSigma^{-1}\UB=\sigmainv\textrm{diag}(1,\ldots,1,w_1,\allowbreak\ldots,w_p),
\]
where $w_i=\lambda\alpha_i/(\sigmainv+\lambda\alpha_i)$ for $i=1,\dots,p$, and the columns of $\UB$ form an orthonormal basis of eigenvectors in $\RR^n$. As elements of the sample space, each of the eigenvectors represent different behaviours over the spatial domain. The first $n-p$ eigenvectors $\UB_{\mathrm{ns}}$ have eigenvalue $\sigmainv$ and are unaffected by smoothing. Therefore, the subspace spanned by these can be considered as the ``non-spatial" part of the sample space. The remaining $p$ eigenvectors $\UB_{\mathrm{sp}}$ represent the spatially smoothed behaviours, where the $i$th eigenvector has eigenvalue $\sigmainv w_i$ with $w_i$ a weight between $0$ and $1$, directly linked to the $i$th smoothing penalty $\alpha_i$. Although, the values of the weights depend on the parameters $\sigmainv$ and $\lambda$, which are usually estimated, the eigenvectors do not. Moreover, irrespective of the parameter estimates, the ordering of the weights is the same as the $\alpha_i$'s, i.e.\ 
\[
0\le w_1\le\cdots\le w_p\le 1.
\]
Thus, spatial eigenvectors with weights close to $1$ correspond to behaviours that are penalised more, i.e.\ that are considered ``less spatial" according to the analysis model. We will therefore refer to behaviours given by eigenvectors with high weights $w_i$ as ``high frequency" and eigenvectors with low weights $w_i$ as ``low frequency". For further details, see Appendix~1.2.

Although spatial frequencies are generally known to play an important role for spatial confounding, there is no single definition in the literature for what is meant by spatial frequencies (see Appendix~1.2). Describing them directly through the eigenvectors of the spatial precision matrix $\BSigma^{-1}$ 
has several advantages:
\begin{itemize}
\item For any choice of analysis model (\ref{eqn:model}), these eigenvectors are well-defined and can be easily computed, even when the dimension $p$ of the spatial effect is smaller than the sample size $n$;
\item The eigenvectors form an orthonormal basis for the $n$-dimensional sample space;
\item The eigenvectors are directly and quantifiably linked to the penalisation (i.e. spatial smoothing) in the chosen analysis model (as set out in Lemma \ref{lem:Sigmainv_eig}).
\end{itemize}
Thus, what is considered high/low frequency spatial behaviour is always well-defined and determined by the spatial structure chosen for the spatial analysis model.

Finally, since the precision matrix $\BSigma^{-1}$ is symmetric positive definite it defines an inner product on $\RR^n$. For the rest of this paper, if $\MB$ is a symmetric positive definite $n\times n$ matrix, we will use the notation $\langle\cdot,\cdot\rangle_{\MB}$ to denote the inner product defined by $\MB$ and $\Vert\cdot\Vert_{\MB}$ the corresponding norm. That is, $\langle\vb,\vb'\rangle_{\MB}=\vb^T\MB\vb'$ and $\Vert\vb\Vert_{\MB}=\sqrt{\langle \vb,\vb\rangle_{\MB}}$ for all $\vb,\vb'\in\RR^n$.

\section{Characterising the impact of spatial confounding}\label{sec:bias}
%\subsection{Bias in the spatial model}
In this section, Proposition~\ref{prop:bias} and Corollary~\ref{cor:bias} provide explicit expressions for the bias of the covariate effect estimate in the analysis model (\ref{eqn:model}) arising from spatial confounding. 
We assume throughout that the true data generating process is that of model \eqref{eqn:dgp} with true covariate effect $\beta$ and true spatial confounder $\zb$, where the latter is assumed to be generated from the spatial structure of model \eqref{eqn:model}. Thus, our results do not rely on model mis-specification; the bias occurs even when the analysis model is perfectly able to reproduce the confounder. Further bias could arise when the analysis model is mis-specified, but this is not our focus here.

\subsection{Bias in the spatial model}
\begin{proposition}\label{prop:bias}
{\em The bias of the estimated covariate effect $\betah$ in model (\ref{eqn:model}) is given by}
\[
\E(\betah)-\beta=\frac{\langle \xb,\zb\rangle_{\BSigma^{-1}}}{\langle\xb,\xb\rangle_{\BSigma^{-1}}}
=\frac{\langle \xb,\zb\rangle_{\BSigma^{-1}}}{\Vert\xb\Vert_{\BSigma^{-1}}\Vert\zb\Vert_{\BSigma^{-1}}}\frac{\Vert\zb\Vert_{\BSigma^{-1}}}{\Vert\xb\Vert_{\BSigma^{-1}}}.
\]
\end{proposition}
Thus, the bias has the intuitive interpretation as the correlation between $\xb$ and $\zb$ times the size of $\zb$ relative to the size of $\xb$ in the metric defined by $\BSigma^{-1}$. That is, in this metric, the bias gets large exactly when the covariate and the confounder are very correlated \emph{and} the confounder is relatively large compared to the covariate. The sign of the bias is the same as the sign of the correlation, i.e.\ the sign of $\langle \xb,\zb\rangle_{\BSigma^{-1}}$. The proof of the proposition is in Appendix~1.3.

Combining Proposition~\ref{prop:bias} with the results of Section~\ref{sec:defs}
we get a better understanding of how the bias behaves. Let $\xxi^{x}=(\xxi^{xT}_{\mathrm{ns}},\xxi^{xT}_{\mathrm{sp}})^T$ denote the coordinates of $\xb$ in the eigenbasis $\UB$ such that $\xb=\UB\xxi^x$, i.e.\ a non-zero entry in $\xxi^x$ means that the corresponding basis vector is included in $\xb$. Similarly, let $\zb=\UB\xxi^z$. Since $\zb$ is spatial, $\xxi^z_{\mathrm{ns}}=\zerob$. It is now straightforward to prove the following result (see Appendix~1.4. for the proof).
\begin{corollary}\label{cor:bias}
Let $\UB$ be the orthonormal eigenbasis which diagonalises $\BSigma^{-1}$, and $\xxi^x=(\xxi^{xT}_{\mathrm{ns}},\xxi^{xT}_{\mathrm{sp}})^T$ and $\xxi^z=(\zerob^T,\xxi^{zT}_{\mathrm{sp}})^T$ the coordinates of $\xb$ and $\zb$ in this basis. The bias
of the estimated covariate effect $\betah$ in model (\ref{eqn:model}) is given by
\[
\E(\betah)-\beta=\frac{\sum_{i=1}^{p}\xi_{\mathrm{sp},i}^x\xi_{\mathrm{sp},i}^z w_i}{\sum_{i=1}^{n-p}(\xi_{\mathrm{ns},i}^x)^2+\sum_{i=1}^{p}(\xi_{\mathrm{sp},i}^x)^2 w_i},
\]
where $w_i=\lambda\alpha_i/(\sigmainv+\lambda\alpha_i)$ for $i=1, \ldots, p$.
\end{corollary}

%\subsection{\textcolor{magenta}{Cause of bias}}
The expression in Corollary~\ref{cor:bias} (consistent with (10) of \citet{schnell2019spectral}) shows that bias in the spatial model is directly linked to spatial smoothing, as without smoothing (i.e.\ if $\alpha_i=0$ for all $i$ or $\lambda=0$), the bias would not arise. In other words, adding spatial random effects to the model (\ref{eqn:model_ns}) adjusts for unmeasured spatial confounders thereby, in principle, eliminating bias, however, bias is reintroduced as a result of spatial smoothing. For any given $\lambda>0$, i.e.\ a fixed overall level of smoothing, we see that bias occurs whenever $\xb$ and $\zb$ share a spatial component for which $w_i\ne 0$. More specifically, any confounded component in $\xb$, i.e.\ with $\xi_{\mathrm{sp},i}^x\xi_{\mathrm{sp},i}^z\ne 0$, contributes to the numerator of the bias. %and the size of the contribution is proportionate to the size of the confounder at that frequency. 
At the same time, all components in $\xb$ increase the denominator, thereby reducing the overall size of the bias.

The expression for the bias is not symmetric in $\xb$ and $\zb$. %While both variables affect the expression, 
%which makes sense as the bias is for the effect of $\xb$ and not $\zb$. 
In particular, it is not affected by components in $\zb$ for which the corresponding component in $\xb$ is 0 (i.e. when $\xi_{\mathrm{sp},i}^z\ne 0$ but $\xi_{\mathrm{sp},i}^x= 0$). Thus, any unmeasured spatial variation in the response variable that is independent of the covariate can be ignored for the purposes of the bias. However, such components may affect the overall fit of the model and the estimates of the parameters $\lambda$ and $\sigma^2$ which affect the weights $w_i$.

\subsection{The role of spatial frequencies}
Due to the differences in the weights $w_i$, different spatial frequencies affect the bias differently. 

Looking at the numerator in Corollary~\ref{cor:bias}, any confounded spatial component of $\xb$ (i.e.\ those with $\xi_{\mathrm{sp},i}^x\ne 0$ and $\xi_{\mathrm{sp},i}^z\ne 0$ ) contribute to the bias, and the larger the weight, the larger the contribution. So confounding at high frequencies (where weights are close to $1$) induce larger bias, while confounding at low frequencies may give a negligible contribution. Confounding at unpenalised spatial frequencies does not lead to bias (as they have weights $0$), and non-spatial components in $\xb$ do not contribute to the numerator either. For all confounded frequencies in $\xb$ with $w_i\ne 0$, the contribution to the bias is proportional to the size $\xi_{\mathrm{sp},i}^z$ of the confounder at that frequency. Contributions to the bias from different eigenvectors can partially cancel each other out as they can have different signs.

From the denominator we see that any unconfounded components of $\xb$, that is, either non-spatial components $\xi_{\mathrm{ns},i}^x\ne 0$ or spatial components with $\xi_{\mathrm{sp},i}^x\ne 0$ and $\xi_{\mathrm{sp},i}^z=0$, will reduce the size of the bias, as they increase the denominator without affecting the numerator.
This reduction will be larger when the weight $w_i$ is large, i.e.\ when the spatial frequency is high, and may be negligible or even $0$ for low frequencies. Unconfounded high frequency spatial components of $\xb$ have a similar effect on the bias as non-spatial components (in line with the intuition that high frequencies are ``less spatial"), i.e.\ they contribute nothing to the numerator and are included with weight close to $1$ in the denominator. %This is in line with intuition as high frequency spatial behaviour is considered ``less spatial", which is why it is given a higher smoothing penalty in the model. 

\subsection{Comparison to the non-spatial model}\label{sec:comparison_ns}
%Now, for comparison, we consider the bias of the covariate effect estimate in the non-spatial model (\ref{eqn:model_ns}). Note that, 
Irrespective of the size of bias, the non-spatial model (\ref{eqn:model_ns}) is not an appropriate analysis model for the data generation process \eqref{eqn:dgp}, as the residual spatial variation in the response data violates the assumptions of the model. Indeed, this is the reason for including spatial random effects. Nevertheless, we consider how the bias in model (\ref{eqn:model_ns}) compares to the spatial model (\ref{eqn:model}) as this has been a focus of spatial confounding research in the past. The following corollary is proved in Appendix~1.8. %as it aids intuition and has also been the focus of much of the research into spatial confounding in the past. The following corollary is proved in the appendix. %) shows that the bias in the non-spatial model is the same as the expression given in Proposition~\ref{prop:bias} but where the inner product and norm are defined by the usual Euclidean metric.

\begin{corollary}\label{cor:bias_ns}
Let $\UB$ be the orthonormal eigenbasis which diagonalises $\BSigma^{-1}$, and $\xxi^x=(\xxi^{xT}_{\mathrm{ns}},\xxi^{xT}_{\mathrm{sp}})^T$ and $\xxi^z=(\zerob^T,\xxi^{zT}_{\mathrm{sp}})^T$ the coordinates of $\xb$ and $\zb$ in this basis.
The bias of the estimated covariate effect $\betah_{\mathrm{ns}}$ in model (\ref{eqn:model_ns}) is given by
\[
\E(\betah_{\mathrm{ns}})-\beta=\frac{\langle \xb,\zb\rangle}{\langle\xb,\xb\rangle}=\frac{\langle \xb,\zb\rangle}{\Vert\xb\Vert\Vert\zb\Vert}\frac{\Vert\zb\Vert}{\Vert\xb\Vert}\\
=\frac{\sum_{i=1}^{p}\xi_{\mathrm{sp},i}^x\xi_{\mathrm{sp},i}^z}{\sum_{i=1}^{n-p}(\xi_{\mathrm{ns},i}^x)^2+\sum_{i=1}^{p}(\xi_{\mathrm{sp},i}^x)^2}.
\]
\end{corollary}

%Corollary~\ref{cor:bias_ns} shows that the bias in the non-spatial model can be interpreted as the (usual) correlation between the covariate and the spatial effect times the size of the spatial effect relative to the size of the covariate. Thus,
Thus, the bias in the non-spatial model differs to that of the spatial model only in that all spatial frequencies have weights $1$ rather than $w_i$. Broadly speaking, bias in the non-spatial model reflects the overall correlation between $\xb$ and $\zb$, whereas bias in the spatial model depends mostly on the correlation at high frequencies. However, as the weights affect both the numerator and denominator of the bias, in general, the relationship between the two models is not straightforward. A detailed computation of the difference and how it is affected by different types of behaviours is given in Appendix~1.4. %Appendix \ref{app:ns}
Although the bias tends to be larger in the non-spatial model than in the spatial model, there are scenarios where it is the other way around. %In particular, if the covariate has little non-spatial or unconfounded high frequency information and, at the same time, confounding is at high frequencies, then the bias in the spatial model may well exceed that of the non-spatial model. }

In the past, the estimate in model \eqref{eqn:model_ns} was often mistakenly assumed to be unbiased and, therefore, a large difference between this and the estimate in \eqref{eqn:model} was seen as a sign of confounding bias in the spatial model (and, conversely, a small difference a sign of no confounding). However, Corollary~\ref{cor:bias_ns} shows that a difference between these estimates tends to arise when the covariate is dominated by low frequencies (as these have weights most different from $1$) and the existence of such a difference does not necessarily mean that there is confounding bias in the spatial model. %e.g.\ a large low frequency confounder is likely to cause significantly larger bias in the non-spatial model than in the spatial model.
Conversely, we see that if the covariate is dominated by high frequencies (with weights close to $1$), the estimates in the two models will be similar, but this does not necessarily mean that they are unbiased.

\subsection{Dependence on the smoothing parameter}
%Corollary~\ref{cor:bias} can also be used to consider the dependence of the bias on the overall level of smoothing in the spatial model, controlled by the parameter $\lambda>0$. 
As bias in the spatial model is caused by smoothing, intuitively, we would expect larger values of the parameter $\lambda$ to lead to more bias when all other components are held fixed. However, as $\lambda$ affects both the numerator and the denominator of the expression in Corollary~\ref{cor:bias}, there can be situations where this is not the case. %Specifically, the dependence of the bias on $\lambda$ is through the weights $w_i(\lambda)$, which all increase from 0 to 1 at different rates. The way these weights are combined in the expression for the bias is determined by the confounding scenario and the parameters of the model. 
In  Appendix~1.5 % Appendix~\ref{app:lambda} 
we provide a detailed analysis which shows that, although the behaviour of the bias varies, %bias for a given $\xb$ and $\zb$ can have different behaviours as a function of $\lambda$, 
when there is sufficient non-spatial or unconfounded high frequency components in $\xb$, the size of the bias broadly increases with $\lambda$ (in line with intuition). The following corollary is also proved in the Appendix~1.6. %a general result on the behaviour in the limits $\lambda\rightarrow 0$ and $\lambda\rightarrow \infty$.
%\textcolor{magenta}{If we're tight on space, we could maybe move the corollary to the appendix and only include the summary of results?}
\begin{corollary}\label{cor:bias_lambda}
Let $\UB$ be the orthonormal eigenbasis which diagonalises $\BSigma^{-1}$, and $\xxi^x=(\xxi^{xT}_{\mathrm{ns}},\xxi^{xT}_{\mathrm{sp}})^T$ and $\xxi^z=(\zerob^T,\xxi^{zT}_{\mathrm{sp}})^T$ the coordinates of $\xb$ and $\zb$ in this basis. The bias of the estimated covariate effect $\betah$ in model (\ref{eqn:model}) has the following limiting behaviour:
\begin{eqnarray*}
\lim_{\lambda\rightarrow 0}\E(\betah)-\beta&=&
\left\{\begin{array}{cr} 
0&\text{if }\xxi_{\mathrm{ns}}^x\ne \zerob\\  \frac{\sum_{i=1}^{p}\xi_{\mathrm{sp},i}^x\xi_{\mathrm{sp},i}^z\alpha_i}{\sum_{i=1}^{p}(\xi_{\mathrm{sp},i}^x)^2\alpha_i}&\text{otherwise}\end{array}\right.\\
\lim_{\lambda\rightarrow\infty}\E(\betah)-\beta&=&\frac{\sum_{\{i|\alpha_i\ne 0\}}\xi_{\mathrm{sp},i}^x\xi_{\mathrm{sp},i}^z}{\sum_{i=1}^{n-p}(\xi_{\mathrm{ns},i}^x)^2+\sum_{\{i|\alpha_i\ne 0\}}(\xi_{\mathrm{sp},i}^x)^2}.
\end{eqnarray*}
\end{corollary}
Thus, in the limit $\lambda\rightarrow \infty$, i.e.\ an entirely smoothed spatial effect, the bias in the spatial model agrees with that of the non-spatial model, except that any contributions from unpenalised spatial components of $\xb$ are $0$.
%This difference arises because if such components exist (these would be the lowest spatial frequencies), then no matter how much the overall smoothing in the spatial model is increased, the unpenalised part of the spatial effect will remain constant. %However, unless there is a large proportion of unpenalised spatial components in $\xb$, we expect the bias in this limit to be similar to that of the non-spatial model.
In the limit $\lambda\rightarrow 0$, i.e.\ without smoothing, the behaviour of the bias depends on whether or not the covariate has non-spatial information. If non-spatial information is present, the bias can be eliminated by setting $\lambda=0$, though this is usually undesirable as it is likely lead to an inferior model fit. Similarly, if the proportion of unconfounded high frequency components in $\xb$ is large, then the size of the bias reduces as $\lambda\rightarrow 0$ and the limit is close to $0$. %, in line with the intuition that unconfounded high frequency components are similar to non-spatial components. 
But in the absence of non-spatial or unconfounded high frequency information, the limit could become large and even exceed the bias of the non-spatial model.

\section{Simulation studies}\label{sec:simulations}
%\subsection{Simulation setup}
%\textcolor{magenta}{Figures: I suggest for Scenarios 1-4 we put all the MSE figures in the appendix (since the conclusion is always just that non-spatial mse is worse than spatial mse). We could then have one plot with the biases in scenarios 1-4 (see my comments at the beginning of scenario 4), one plot for scenario 5, and one plot (with both bias and mse) for scenario 6.}

In the following sections we simulate different confounding scenarios to study the behaviour of the bias of the covariate effect estimate $\betah$ in model \eqref{eqn:model} and $\betah_\mathrm{ns}$ in model \eqref{eqn:model_ns}. %Specifically, we consider: (i) confounding at high frequencies, (ii) confounding at  low frequencies, (iii) confounding at high frequencies but low overall correlation, (iv) different proportions of non-spatial information in $\xb$, and (v) increasing values of the smoothing parameter $\lambda$. 
For each scenario we generate 100 independent replicates of covariate data $\xb$ and response data $\yb$, observed at $n = 1000$ randomly selected locations in the spatial domain $[0,1]\times[0,1] \subset \mathbb{R}^2$ such that
$$\yb= \beta \xb + \zb + \eepsilon^y, \qquad \xb = \zb^x+ \eepsilon^x$$
where $\beta\in\RR$, $\eepsilon^y \sim N(\zerob, \sigma^2 \IB)$, $\eepsilon^x \sim N(\zerob, \sigma^2_x \IB)$ for $\sigma, \sigma_x>0$. In this specification, $\zb$ and $\zb^x$ are constructed as linear combinations of two spatial fields $\zb_{\mathrm{sp,l}}$ and $\zb_{\mathrm{sp,h}}$, which are low and high frequency, respectively, i.e.\
$$\zb = \xi_{\mathrm{sp, l}}^z \zb_{\mathrm{sp,l}} + \xi_{\mathrm{sp, h}}^z \zb_{\mathrm{sp,h}},\qquad \zb^x = \xi_{\mathrm{sp, l}}^x \zb_{\mathrm{sp,l}} + \xi_{\mathrm{sp, h}}^x \zb_{\mathrm{sp,h}}$$
for $\xi_{\mathrm{sp, l}}^z, \xi_{\mathrm{sp, h}}^z, \xi_{\mathrm{sp, l}}^x, \xi_{\mathrm{sp, h}}^x \in\RR$, while $\eepsilon^x$ is spatially unstructured and represents the non-spatial component of $\xb$. Unless stated otherwise, we let $\beta = 0.5$, $\sigma_x = \sigma = 1$.

For the regressions in this section, we represent the spatial effect as a thin plate regression spline (TPRS), a computationally efficient reduced rank version of a thin plate spline \citep{wood2003thin}.
The data generating process matches the analysing spatial model which is therefore correctly specified.
In Appendix~4, %Appendix~\ref{app:simulation_gps}
we repeat the simulation where the data generating process is a Gaussian process, so in addition to the spatial confounding bias, there may be further mis-specification bias. In order to generate $\zb_{\mathrm{sp,l}}$ and $\zb_{\mathrm{sp,h}}$, we start by considering a mean-zero Gaussian process $\gammab$ with exponential covariance structure following $C(h) = \exp(-h/\kappa)$ such data $h = \lVert \ssb - \ssb^\prime \rVert$ for $\ssb, \ssb^\prime \in [0,1]\times[0,1]$, which is assumed for simplicity to have variance of 1.
We set $\kappa = 0.1$ corresponding to a spatial range of approximately 0.3.
We fit thin plate regression splines to the generated $\gammab$ with 10 and 800 basis functions, respectively, and use the corresponding fitted values as $\zb_{\mathrm{sp,l}}$ and $\zb_{\mathrm{sp,h}}$.
Thin plate regression spline models are implemented in the R-package mgcv. We use this implementation with generalised cross-validation as the smoothness selection criterion to compare the results of models fitted to simulated data for which we know the true underlying covariate and spatial dependence.
We consider 600 basis functions for the thin plate regression splines in the data analysing model.

\subsection{Scenario~1: Confounding at high frequencies}
From Corollary~\ref{cor:bias} we expect confounding at high frequencies, i.e.\ $\xi_{\mathrm{sp, h}}^x \xi_{\mathrm{sp, h}}^z \ne 0$, to lead to bias in the spatial model. The larger the value of $\xi_{\mathrm{sp, h}}^x \xi_{\mathrm{sp, h}}^z $, the larger the numerator in the bias. In fact, keeping everything else fixed, the bias increases linearly with $\xi_{\mathrm{sp, h}}^z$ and can therefore become arbitrarily large. From Corollary~\ref{cor:bias_ns} we expect the same behaviour in the non-spatial model, as the bias has the same expression as that of the spatial model, except that each weight $w_i$ is replaced by $1$.
Thus, we consider $(\xi_{\mathrm{sp, l}}^x, \xi_{\mathrm{sp, h}}^x) = (1,1)$, $\xi_{\mathrm{sp, l}}^z = 0$ and $\xi_{\mathrm{sp, h}}^z \in \{0.2, 0.5, 1\}$.

\begin{figure}
    \centering
    \begin{minipage}{1\textwidth}
        \centering
        \includegraphics[width=1\textwidth]{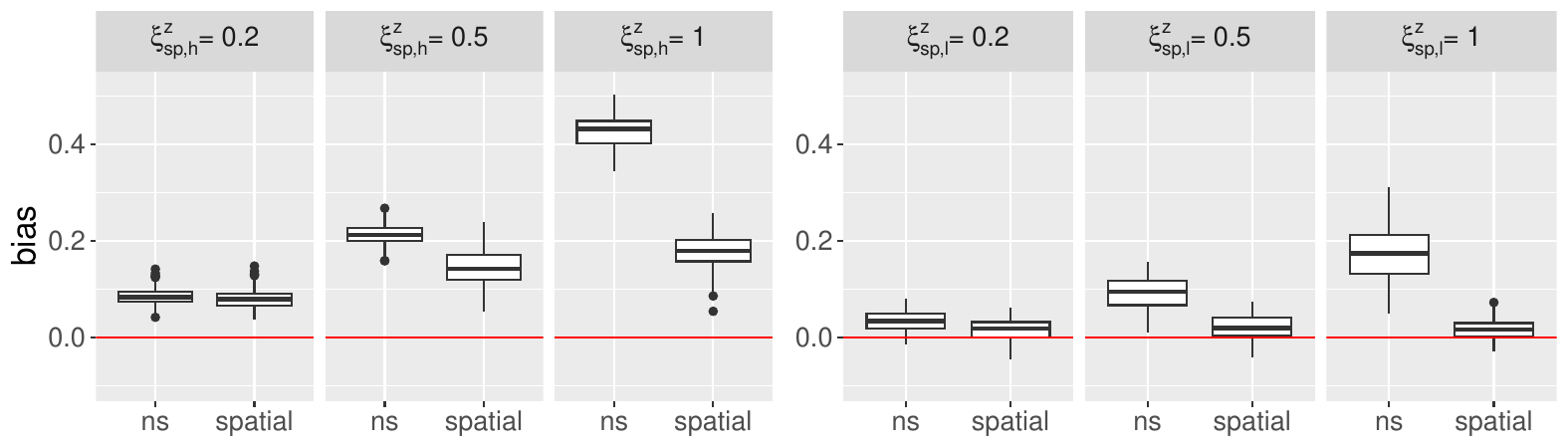} 
    \end{minipage}
    
    \begin{minipage}{1\textwidth}
        \centering
    \includegraphics[width=1\textwidth]{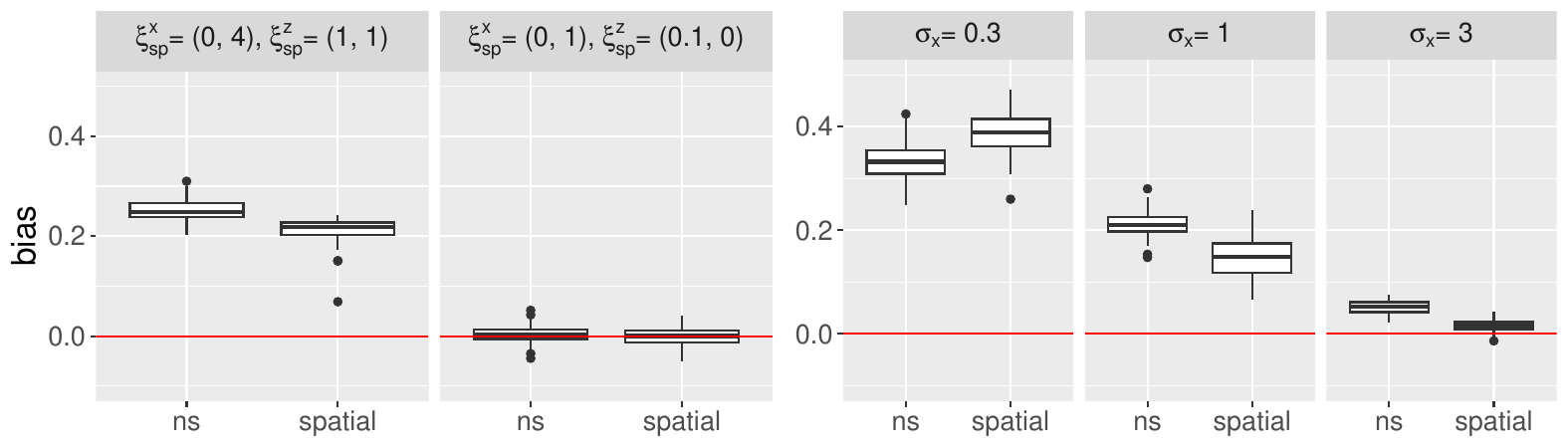}
    \end{minipage}
    \caption{Bias of $\betah$ and $\betah_\mathrm{ns}$ in the spatial and non-spatial (ns) models under Scenario 1 (top, left),  Scenario~2 (top, right), Scenario~3 (bottom, left),  Scenario~4 (bottom, right). The true value of $\beta$ is 0.5.} 
    \label{fig:s1}
\end{figure}

As expected, Figure~\ref{fig:s1} (top left) shows significant bias in both models under this scenario, increasing linearly in size with $\xi_{\mathrm{sp, h}}^z$. 
The mean squared error (MSE) of fitted values compared to the true expectation of $\yb$ is shown in Appendix~3.1 % Appendix~\label{app:mse} 
and is always lower for the spatial model, indicating a better fit, which is not surprising as the spatial model is correctly specified, whereas the non-spatial model is mis-specified.

\subsection{Scenario~2: Confounding at low frequencies}
We reconsider Scenario~1 but with high and low frequencies swapped, i.e.\ $(\xi_{\mathrm{sp, l}}^x, \xi_{\mathrm{sp, h}}^x) = (1, 1)$, $\xi_{\mathrm{sp, h}}^z = 0$ and $\xi_{\mathrm{sp, l}}^z\in \{0.2, 0.5, 1 \}$.
Since confounding is now at low frequencies, i.e.\ $\xi_{\mathrm{sp, l}}^x \xi_{\mathrm{sp, l}}^z \ne 0$, contributions to the numerator of the bias in the spatial model are multiplied by low weights $w_i$, while the denominator of the bias is still relatively large due to the non-spatial information in $\xb$. Therefore we expect the bias to be small. This is confirmed in the results shown in Figure~\ref{fig:s1} (top right) where the bias in the spatial model remains close to zero, even when $\xi_{\mathrm{sp, l}}^z$ is increased.
In contrast, swapping frequencies makes no difference to the non-spatial model as all frequencies have equal weight and, indeed, the behaviour for this model is the same as in Scenario 1. Appendix~3.1 %Appendix~\ref{app:mse} 
confirms, as before, that the spatial model is superior in terms of fit.

\subsection{Scenario~3: Comparison to the non-spatial model}%Unconfounded low frequencies}
In Section~\ref{sec:comparison_ns} we noted that difference or similarity between the estimates in the spatial and non-spatial models cannot in itself be used to diagnose the existence or otherwise of bias in the spatial model. This is illustrated in Scenarios~1 and 2 above, as in both of these scenarios the estimates in the two models differ but, in Scenario~1, the spatial model is biased whereas, in Scenario~2, it is not. To illustrate this point further, here we simulate two scenarios in which the estimates in the two models are similar, but in one scenario both models are biased whereas in the other they are both largely unbiased. Specifically, in order to generate $\zb_{\mathrm{sp,h}}$ and $\zb_{\mathrm{sp,l}}$, we fit a reparametrised TPRS spatial effect (see Appendix~2) %Appendix~\ref{app:reparam}) 
to the Gaussian process with $\kappa = 0.1$  where the corresponding 75 highest frequencies are used to generate $\zb_{\mathrm{sp,h}}$ and the remaining frequencies to generate $\zb_{\mathrm{sp,l}}$. We consider: (1) biased scenario with $(\xi_{\mathrm{sp, l}}^x, \xi_{\mathrm{sp, h}}^x) = (0, 4)$ and $(\xi_{\mathrm{sp, l}}^z, \xi_{\mathrm{sp, h}}^z) = (1, 1)$; (2) unbiased scenario with $(\xi_{\mathrm{sp, l}}^x, \xi_{\mathrm{sp, h}}^x) = (0, 0.1)$ and $(\xi_{\mathrm{sp, l}}^z, \xi_{\mathrm{sp, h}}^z) = (1, 0)$.  We use $n$ basis functions in the the data analysis model. In the first scenario, $\xb$ has only high frequencies which are all confounded, so the spatial model is biased with weights $w_i$ close to $1$, making the bias similar to that of the non-spatial model. In the second scenario, there are no overlapping frequencies between $\xb$ and $\zb$ so both models are unbiased. The results are shown in Figure~\ref{fig:s1} (bottom, left). %The biased scenario is based on the fact that if $\xb$ only has high frequencies, which are the most heavily smoothed, and these are also confounded with $\zb$, then the bias should be similar. In the unbiased scenario, there are no shared frequencies and additionally $\zb$ has a small weight thus reducing model mis-specification bias in the non-spatial model. 

Section~\ref{sec:comparison_ns} also showed that while the bias in the non-spatial model tends to exceed that of the spatial model, it could also be the other way around. In both Scenarios~1 and 2, the bias in the non-spatial model always stays above the spatial model bias. However, in Appendix~3.2, %Appendix~\ref{app:s3}
we repeat Scenario~1 with a larger proportion of unconfounded low frequency components in $\xb$ so that the correlation between $\xb$ and $\zb$ is low overall, but high at high frequencies. As expected from our analysis, in this latter scenario, the bias in the spatial model exceeds that of the non-spatial model. 

\subsection{Scenario~4: Dependence on non-spatial information}
%\textcolor{magenta}{\emph{Maybe leave out $\sigma_x=2$ here as there are then only three sigmas to plot. Then for Scenario~4, don't repeat Scenario~1. Then we could amalgamate Figures 3 and 4 to take up less space - mostly in the figures but also a little bit in the writing.}}
We now study the effect on the bias of non-spatial information in the covariate $\xb$ by increasing the parameter $\sigma_x$. As the coefficient $\xi_{\mathrm{ns}}^x$ appears only in the denominator of the bias in both the spatial and non-spatial models, we expect bias in both models to reduce.
Specifically, we consider $\sigma_x \in \{0.3, 1, 3 \}$. Other parameters remain constant at $(\xi_{\mathrm{sp, l}}^x, \xi_{\mathrm{sp, h}}^x) = (1, 1)$, $(\xi_{\mathrm{sp, l}}^z, \xi_{\mathrm{sp, h}}^z) = (0,0.5)$.

Figure~\ref{fig:s1} (bottom, right) shows that, as expected, the bias decreases for both models when $\sigma_x$ is increased, and the case $\sigma_x=3$ illustrates that, for a sufficiently large non-spatial component of $\xb$, the bias in the spatial model becomes negligible. As before, the MSE of fitted values is always larger for the non-spatial model.
In Section~\ref{sec:comparison_ns}, we saw that bias in the non-spatial model tends to exceed that of the spatial model, but if the covariate has little non-spatial information and confounding takes place at high frequencies, then it may be the other way around. This behaviour is observed in Figure~\ref{fig:s1} where the bias in the non-spatial model is higher than that of the spatial model, except in the case $\sigma_x=0.3$, i.e\ where $\xb$ has the smallest amount of non-spatial information.

% \begin{figure}
%     \centering
%     \begin{minipage}{0.8\textwidth}
%         \centering
%         \includegraphics[width=1\textwidth]{figures/simulation3_results_option3.pdf}
%     \end{minipage}
    
%     \begin{minipage}{0.8\textwidth}
%         \centering
%     \includegraphics[width=1\textwidth]{figures/simulation3_mse_results_option3.pdf}
%     \end{minipage}
%     \caption{Bias of $\betah$ and $\betah_\mathrm{ns}$ (top) and mean squared error of fitted values (bottom) in the spatial and non-spatial (ns) models under Scenario~4. The true value of $\beta$ is 0.5.}
%     \label{fig:s3}
% \end{figure}

\subsection{Scenario~5: Dependence on the smoothing parameter}
As detailed in Section~\ref{sec:bias}, the expression for the bias in the spatial model has a relatively complex dependency on the smoothing parameter $\lambda$. However, when there are sufficient non-spatial or unconfounded high frequency components in $\xb$, we expect the bias to broadly increase from zero and, assuming there are not many unpenalised spatial components in $\xb$, eventually approach the bias in the corresponding non-spatial model. 
%generally, the bias will be close to zero as $\lambda \rightarrow 0$, broadly increase as a function of $\lambda$ and (assuming there are not many unpenalised spatial components in $\xb$) approach the bias in the corresponding non-spatial model as $\lambda \rightarrow \infty$.
%However, and as shown in Scenario~3, it can also exceed that of the non-spatial model.
Here, we simulate this type of data under three data generation processes, chosen to broadcast a range of behaviours. 
The simulated data has confounding at high frequencies, unconfoundedness at low frequencies and varying amounts of non-spatial information, specifically, $\sigma_x \in \{0.3, 1, 2 \}$. 
We use 10 basis functions to generate $\zb_{\mathrm{sp, l}}$ and 800  basis functions for $\zb_{\mathrm{sp, h}}$.
Moreover, we let $(\xi_{\mathrm{sp, l}}^x, \xi_{\mathrm{sp, h}}^x) = (1, 1)$, $(\xi_{\mathrm{sp, l}}^z, \xi_{\mathrm{sp, h}}^z) = (0, 0.5)$ and fix $\lambda$ at different values. We consider 20 replicates and compute the average spatial and non-spatial bias for those replicates using 800 basis functions in the analysing spatial model.

%While for higher values of $\sigma_x$ the spatial bias should mostly stay below the non-spatial bias, for lower values of $\sigma_x$ this may not be the case. %We expect the spatial bias to approach the non-spatial bias for large values of $\lambda$ in every case.

\begin{figure}
    \centering
    \includegraphics[width = 1\textwidth]{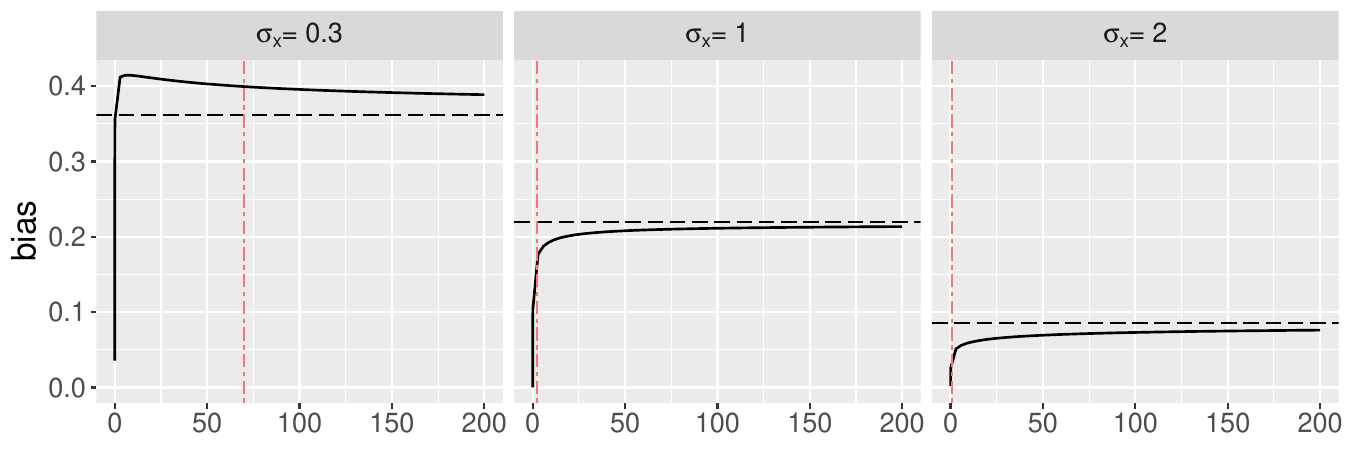}
    \caption{Bias of $\betah$ (solid black) and $\betah_\mathrm{ns}$ (dashed black) for increasing $\lambda$ and different $\sigma_x$. The true value of $\beta$ is 0.5. The magenta dashed line shows the median $\hat{\lambda}$ estimated by generalised  cross-validation.} %The bias associated with each  $\hat{\lambda}$ is 0.4037, 0.1804 and 0.0362, from left to right.}
    \label{fig:s4}
\end{figure}

Figure~\ref{fig:s4} shows that for $\sigma_x = 0.3$, the spatial bias 
increases from 0, crosses that of the non-spatial model and stays above it, until they meet again for large $\lambda$.
As expected, increasing non-spatial information in $\xb$ decreases bias in general, and for $\sigma_x = 1$ and $\sigma_x=2$, the bias of the spatial model stays below that of the non-spatial model.
%In Scenarios~1 -- 4, the smoothing parameter was estimated using the generalised cross-validation criterion. 
For each data generating process in Figure~\ref{fig:s4}, we also show the median estimated smoothing parameter $\hat{\lambda}$ using the generalised cross-validation criterion. At the estimated values of $\lambda$ the bias is 0.4037, 0.1804 and
0.0362, from left to right. For $\sigma_x = 0.3$, the generalised cross-validation estimate is in a relatively flat area of the curve where the bias in the spatial model exceeds that of the non-spatial model. For both of the other values of $\sigma_x$, although the spatial model bias is bounded by the non-spatial model bias, the estimate $\hat{\lambda}$ is at the steep part of the curve, which means that small changes in $\hat{\lambda}$ can lead to large changes in the bias. 
 
\section{Bias adjustment}\label{sec:adj}%\label{sec:understanding}
%The expressions derived in Section~\ref{sec:bias} provide an intuitive and quantifiable way to assess the impact of spatial confounding in the general formulation of a spatial mixed model. 

%In the following sections, we use our results from Section~\ref{sec:bias} as a unified theoretical framework for understanding and generalising the main results on spatial confounding to date. Rather than an in-depth analysis of specific methods, this is intended as an overview that illustrates how our results can be applied in practice, not only for gaining intuition, but also for understanding how different aspects of the spatial confounding literature relate to each other.

As the expressions in Section~\ref{sec:bias} involve the unknown confounder $\zb$, in practical applications, we cannot simply compute the bias to assess the impact of confounding. Moreover, as illustrated in Section~\ref{sec:comparison_ns} and the Scenario~3 simulations above, the difference between the estimates in the spatial and non-spatial models is a poor diagnostic for the existence or otherwise of bias.
%In the past, the difference between the estimates in the spatial and non-spatial models has often been used to identify confounding bias. However, as illustrated in Section~\ref{sec:comparison_ns} and the Scenario~3 simulations above, this difference is not directly related to the size or occurrence of bias. 
Here, instead we propose a general framework for bias assessment and adjustment based on our results. %Here, we use our results to develop a general framework for putting into practice different methods for bias adjustment. 
That is, for a given data set and choice of spatial analysis model, we propose methods for estimating the bias under different assumptions for the data generation scenario. %as well as clarifying the assumptions associated with each method. %However, the expressions clarify which scenarios are more/less likely to lead to bias and help us understand the mathematical implications of applying different methods.
One assumption that is not usually highlighted in the spatial confounding literature is whether the covariate $\xb$ has non-spatial information. %, i.e.\ whether $\xxi^x_{\mathrm{ns}}$ is non-zero. %In most references, this assumption is only implicit and needs to be inferred from the setup of a theoretical analysis or the way in which data are generated in a simulation. 
However, as illustrated in Sections~\ref{sec:bias} and \ref{sec:simulations}, %by the comparison to the non-spatial model and the dependence of the bias on the smoothing parameter $\lambda$, 
the bias may exhibit fundamentally different behaviour depending on whether or not this assumption holds. As outlined below, this assumption is also important in the context of bias adjustment. %method used for estimating or adjusting for bias. %Since our expressions for the bias in Section \ref{sec:bias} include $\xxi^x_{\mathrm{ns}}$, they can be used to consider both the case where $\xxi^x_{\mathrm{ns}}$ is non-zero and where it is not.

\subsection{Case 1: $\xb$ has non-spatial information} 
If the covariate $\xb$ has non-spatial information, i.e.\ $\xxi^x_{\mathrm{ns}}\ne\zerob$, it means that, while the locations at which the data are collected induce some spatial dependency structure in the covariate data, there is additional variation that cannot be explained by the spatial information alone. 
%Assume that the covariate $\xb$ has non-spatial information, i.e.\ $\xxi^x_{\mathrm{ns}}\ne\zerob$. This means that, while the locations at which the data are collected induce some spatial dependency structure in the covariate, there is some additional variation in the covariate data that cannot be explained by the spatial information alone. 
In this case, the model matrix for (\ref{eqn:model}) has full rank and the model is identifiable. Thus, the model has information to distinguish the effect of $\xb$ from that of $\zb$, but bias in the covariate effect estimate arises as a result of distortion caused by spatial smoothing. %Without smoothing, i.e.\ if $\lambda=0$, the covariate effect estimate is the ordinary least squares estimate, which for a correctly specified model is unbiased. 
%Smoothing is applied here in order to %improve model fit, i.e.\ to reflect the spatial covariance structure and avoid overfitting the spatial effect, however, while it is only applied to the spatial effect, it also induces bias in the effect estimate of the {\em covariate}. 
If our only objective was to avoid the bias, we could simply remove or reduce the level of smoothing, indeed, Corollary~\ref{cor:bias_lambda} shows that the bias becomes zero as $\lambda\rightarrow 0$. However, undersmoothing is undesirable from the point of view of model fit, as smoothing is applied to 
reflect the spatial covariance structure and avoid overfitting the spatial effect.

Spatial+ \citep{dupont2021spatial+} %and the geoadditive structural equations model \citep{thaden2018structural} are two methods 
is a method for bias adjustment which utilises the fact that the non-spatial component $\xb_{\mathrm{ns}}=\UB_{\mathrm{ns}}\xxi^x_{\mathrm{ns}}$ of the covariate is unconfounded and can therefore be used to identify the covariate effect. Although originally introduced for models in which the spatial effect is a thin plate spline, the method can be formulated for any spatial analysis model as the following two-step procedure.
\begin{description}
    \item[(i)] Fit the model (\ref{eqn:model}) with no covariates to $\xb$ to obtain the decomposition $\xb=\xb_{\mathrm{sp}}+\rb^x$ where $\xb_{\mathrm{sp}}$ are the fitted values (estimating the spatial part of $\xb$) and $\rb^x$ the residuals.
    \item[(ii)] Estimate the covariate effect $\beta$ by fitting the model (\ref{eqn:model}) but with $\xb$ replaced by $\rb^x$.
\end{description}
From Proposition~\ref{prop:bias}, the bias in spatial+ is
$\langle \rb^x,\zb\rangle_{\BSigma^{-1}}/\langle \rb^x,\rb^x\rangle_{\BSigma^{-1}}$. The residuals $\rb^x$ estimate the component $\xb_{\mathrm{ns}}$ which lies in the non-spatial part of the sample space. As this subspace is orthogonal to the spatial part containing $\zb$, this shows why we can expect the bias to be eliminated. Indeed, even though $\rb^x$ may contain a small proportion of potentially confounded spatial components, its non-spatial part is likely to dominate so that bias is negligible (as seen in the Scenario 4 simulations).

\subsection{Case 2: $\xb$ is fully spatial}\label{sec:case_2}
Sometimes $\xb$ has no non-spatial components, i.e.\ $\xxi^x_{\mathrm{ns}}=\zerob$. For example, $\xb$ could be entirely generated from a smooth spatial process, or the dimension $p$ of the spatial effect in (\ref{eqn:model}) could be as large as the sample size $n$ (such as in a regional spatial model with one data measurement per region). The covariate vector is then contained in the space spanned by the spatial basis vectors and model \eqref{eqn:model} is unidentifiable. Here, the smoothing penalty not only improves the fit but acts as a regulariser. The model cannot be fitted without smoothing, %, i.e.\ if $\lambda=0$. 
and the apportionment of the covariate effect between the covariate and spatial terms of the model is directly related to how the smoothing is applied. %the estimate of $\lambda>0$.
%Any fit of the model corresponds to a choice of $\lambda>0$, and the split of the covariate effect between the covariate and spatial terms of the model is directly related to this choice. %, which may not correspond to the true split. 
Our expression for the bias still holds, %with the same conclusions about which situations cause most bias. But 
however, unlike Case 1 above where the non-spatial component $\xb_{\mathrm{ns}}$ is known to be unconfounded, we have no information of the confounding scenario and methods like spatial+ may not be effective. The behaviour of the bias will also generally be harder to predict. In particular, even if confounding is at low frequencies (making the numerator of the bias small), the bias could still be significant as the absence of non-spatial components in $\xb$ means that the denominator could also potentially be very small, especially when the covariate is dominated by low frequency behaviours.
%we no longer have the result that the bias approaches zero as $\lambda\rightarrow 0$ and, contrary to intuition, the bias could even increase as smoothing is reduced.

As shown in Sections~\ref{sec:bias} and \ref{sec:simulations}, spatial frequencies play a key role for the bias, and techniques such as Fourier and wavelet analysis are useful for understanding the behaviours of the observed variables $\yb$ and $\xb$ (as suggested, for example, in \citet{keller2020selecting}). However, our expressions show that it is the frequencies shared by $\xb$ and $\zb$ that cause bias, and %as $\zb$ is unknown, we cannot deduce from the data alone 
the data has no information on how such frequencies are split between the two variables. Therefore, without further (untestable) assumptions for identifiability, we cannot estimate the bias.

In the case where $\xb$ and $\zb$ are realisations of two stationary Gaussian spatial processes, \citet{guan2020spectral} use a Fourier transform to describe spatial frequencies and show that an assumption of unconfoundedness at high frequencies can be used to obtain identifiability. As noted in \citet{guan2020spectral}, this assumption seems natural as the highest frequency spatial variation in a variable is more likely to be characteristic to that particular variable and therefore not confounded with other spatial variables. %\citet{guan2020spectral} suggest using a spline-based method in the Fourier frequency domain to obtain an estimate of the appropriate threshold. Based on this they then estimate the confounder $\zb$.
Corollary \ref{cor:bias} shows that this idea extends to our framework in which spatial frequencies are defined directly in terms of the spatial analysis model (\ref{eqn:model}) as described in Section~\ref{sec:precision}. Unconfoundedness at the highest frequencies then corresponds to the assumption that $\xi^z_{\mathrm{sp},i}= 0$ for frequencies above a certain threshold. Thus, under this assumption, the frequency components of $\xb$ above the threshold can be used to obtain identifiability. 

We note that the frequencies used to obtain identifiability need not be the highest spatial frequencies. An example of this is the scenario of \cite{bolin2024spatial} who consider the asymptotic behaviour of the covariate effect estimate in an analysis model of the form (\ref{eqn:model}) when the true data generation process is given by
\[
\yb = \beta\mathcal{S}\xb+\eepsilon,\quad \eepsilon\sim N(\boldsymbol{0},\sigma^2\IB)
\]
where $\mathcal{S}$ is a smoothing operator, for example, $\mathcal{S}\xb$ could be the spatial frequencies of the covariate below a threshold $k$ and $\xb-\mathcal{S}\xb$ the remaining (higher) frequencies. Here, there is no spatial confounder as such (in the form of a missing or unmeasured spatial variable), but bias arises due to mis-specification of the covariate in the analysis model. However, since $\beta\mathcal{S}\xb = \beta\xb-\beta(\xb-\mathcal{S}\xb)$, the true data generation process can be written in the form of (\ref{eqn:dgp}) with $\zb=-\beta(\xb-\mathcal{S}\xb)$, and we can therefore use our framework to assess the bias in this case. From the expression for $\zb$ we see that ``confounding" is at the higher frequencies $\xb-\mathcal{S}\xb$ of the covariate, and the lower frequencies $\mathcal{S}\xb$ of the covariate are unconfounded. Thus, by Corollary \ref{cor:bias}, the bias is given by $\frac{-\beta\sum_{i=k+1}^n(\xi^x_{\textrm{sp},i})^2w_i}{\sum_{i=1}^n (\xi^x_{\textrm{sp},i})^2w_i}$. As the sample size $n$ goes to infinity, we expect the higher frequencies to increasingly dominate $\xb$ and therefore the bias to approach $-\beta$, consistent with the results of \cite{bolin2024spatial}.

\subsection{Capped spatial+}\label{sec:capped_splus}
Based on our analysis in Section~\ref{sec:case_2}, we present an adapted version of spatial+ which can be used to assess the bias in the more challenging Case 2 scenario. As previously noted, unconfounded high frequencies in $\xb$ have a similar effect on the bias as non-spatial information. Therefore, assuming the $k$ highest frequencies are unconfounded, a natural way to adapt spatial+ is the following two-step procedure. Let $\ub_{\mathrm{sp},1},\ldots, \ub_{\mathrm{sp},n}$ denote the spatial eigenvectors in $\RR^n$.
\begin{description}
\item[(i)] Decompose the covariate as $\xb = \xb_{\mathrm{sp}}^k+\rb^k$ where $\xb_{\mathrm{sp}}^k=\sum_{i=1}^{n-k}\xi^x_{\mathrm{sp},i}\ub_{\mathrm{sp},i}$ (the lowest $n-k$ frequencies) and $\rb^k=\sum_{i=n-k+1}^{n}\xi^x_{\mathrm{sp},i}\ub_{\mathrm{sp},i}$ (the highest $k$ frequencies).
\item[(ii)] Estimate the covariate effect $\beta$ by fitting the model \eqref{eqn:model} but with $\xb$ replaced by $\rb^k$. 
\end{description}
Proposition~\ref{prop:bias} shows that the bias of the corresponding covariate effect estimate is $\langle \rb^k,\zb\rangle_{\BSigma^{-1}}/\langle \rb^k,\rb^k\rangle_{\BSigma^{-1}}$, which, as $\rb^k$ is assumed to be unconfounded, should therefore, in theory, be $0$. However, as the spatial eigenvectors span the whole of $\RR^n$, $\rb^k$ and $\zb$ no longer lie in orthogonal subspaces. As the component $\rb^k$ can also be modelled as part of the highest frequencies of the spatial effect in \eqref{eqn:model}, the estimate of $\beta$ obtained in this way is highly sensitive to inaccuracies at these frequencies of the estimated spatial effect. Such inaccuracies easily arise in practice as the estimation is based on overall spatial smoothing, driven by the fitted values rather than the behaviour at individual spatial frequencies. In order to avoid such sensitivity, we adapt the model matrix of the regression in (ii) by removing the columns corresponding to the frequencies above $k$. This more explicitly reflects the assumption of unconfoundedness at these frequencies but allows the spatial effect to continue to adjust for confounders at frequencies below the cap. We call the estimate of $\beta$ obtained from (i) and the adapted (ii) the capped spatial+ estimate. By construction, if the assumption of unconfoundedness above the cap is true, the bias for this estimate is $0$. 

In practice, %even if the assumption of unconfoundedness at frequencies above a cap is true, 
the true value of the cap $k$ is unlikely to be known, and it may therefore be mis-specified at a value $k'\ne k$. If $k'<k$, $\rb^{k'} = \sum_{i=n-k'+1}^n\xi^x_{\mathrm{sp},i}\ub_{\mathrm{sp},i}$ is simply a smaller component of $\rb^k$ which is still orthogonal to the spatial part of the adjusted spatial model. Therefore, the bias should still be eliminated. However, if $k'>k$, then $\rb^{k'}$ contains a potentially confounded component of $\xb$, namely, the $k'-k$ highest frequencies of $\xb_{\mathrm{sp}}^k$, and by Corollary~\ref{cor:bias} the bias becomes $\sum_{i=n-k'+1}^{n-k}\xi^x_{\mathrm{sp},i}\xi^z_{\mathrm{sp},i}w_i/\sum_{i=n-k'+1}^n(\xi^x_{\mathrm{sp},i})^2w_i$. For comparison, the bias in the original spatial model (\ref{eqn:model}) in this scenario is $\sum_{i=1}^{n-k}\xi^x_{\mathrm{sp},i}\xi^z_{\mathrm{sp},i}w_i/\sum_{i=1}^n(\xi^x_{\mathrm{sp},i})^2w_i$. Thus, the numerator for the spatial model includes additional confounded frequencies contributing to the bias, but at the same time the denominator is also larger. Therefore, in this case, the capped spatial+ estimate may not reduce and could even increase the size of the bias of the spatial model.

For a given data set in the Case 2 scenario of Section~\ref{sec:case_2}, under the assumption of unconfoundedness at the highest frequencies, the bias can therefore be assessed using capped spatial+ with a range of caps $k'$. Starting at small values of $k'$, the corresponding estimate of $\beta$ should be unbiased and therefore stay stable as we increase $k'$, but when $k'$ reaches the true cap $k$, the estimate changes due to the introduction of confounding bias. Thus, if we observe stable capped spatial+ estimates for caps below a certain value, these estimates can be expected to be unbiased. 

Finally, in Section~\ref{sec:case_2} we noted that the assumption of unconfoundedness does not have to be limited to the highest frequencies of the spatial effect. If we believe the unconfounded behaviour of $\xb$ is in a different frequency range, we can easily adapt the capped spatial+ method by replacing the component $\rb^k$ by the part of $\xb$ that is believed to be unconfounded. In the scenario of \cite{bolin2024spatial} described at the end of Section~\ref{sec:case_2}, %where the unconfounded part is believed to be the lowest frequencies $\mathcal{S}\xb$ of the covariate, 
we could use the $k$ lowest frequencies of $\xb$ as the component $\rb^k$ under the assumption that these are contained in the unconfounded low frequency component $\mathcal{S}\xb$ of the covariate. This is therefore also consistent with the suggestion in \cite{bolin2024spatial} to replace $\xb$ in the analysis model (\ref{eqn:model}) by a smoothed (i.e.\ lower frequency) version $\hat{\mathcal{S}}\xb$ of the covariate.

%Finally, we note that the assumption of unconfoundedness does not have to be limited to the highest frequencies of the spatial effect. If we believe the unconfounded behaviour of $\xb$ is in a different frequency range, we can easily adapt the capped spatial+ method by replacing the component $\rb^k$ by the part of $\xb$ that is believed to be unconfounded.

\subsection{Simulations for capped spatial+}\label{sec:sim_capped}
In this section, we use simulations to illustrate the ideas behind the capped spatial+ method proposed in Section~\ref{sec:capped_splus}. % for the Case 2 scenario of Section~\ref{sec:case_2}. 
For a fixed value of the cap $k$ we simulate data in which the covariate $\xb$ is fully spatial but the $k$ highest frequencies are unconfounded. We then consider the bias in the spatial model (\ref{eqn:model}) as well as the capped spatial+ estimates for different values of the $\mathrm{cap}$. 

The data is generated such that:
$$\yb= \beta \xb + \zb+ \eepsilon^y,\quad \xb = \zb^x_\mathrm{sp, low} + \zb^x_\mathrm{sp, medium} + a\ \zb^x_\mathrm{sp, high},\quad \zb = 
 \zb^z_\mathrm{sp, low} + \zb^x_\mathrm{sp, medium}  $$
where $\eepsilon^y \sim N(\zerob, \sigma^2 \IB)$ and $a>0$ controls the size of the unconfounded high frequency spatial component in $\xb$.
In order to generate $\zb^x_\mathrm{sp, low}$,  $\zb^x_\mathrm{sp, high}$ and $\zb^x_\mathrm{sp, medium}$, we fit a thin plate spline to a realization of a Gaussian process with $\kappa = 0.5/3$ (see beginning of the section),  where the penalty and design matrix have now been reparameterised such that they only consider the $100$ lowest, the $k$ highest and all medium frequencies, respectively; $\zb^z_\mathrm{sp, low}$ is generated similarly to Section~\ref{sec:simulations} but it is fitted to a new realization of a Gaussian process with $\kappa = 0.5/3$ using only the $100$ lowest basis functions.
We consider $k\in \{5, 10, 20\}$ and $a\in \{1, 2\}$ in the data generating process; $\beta = 1$, $\sigma =  0.1$, $n = 1000$, $N=50$. 
The reparametrisation is explained in Appendix~2. %\ref{app:reparam}.
 
In the data analysis model, the capped spatial+ model uses a reparameterised $\BBstilde$ and $\bbetastilde$ in the design and penalty matrix for the spatial effects in the first and second stage equation of spatial+ with $ \mathrm{cap} \in \{ 5, 10, 15, 20, 25, 30, 40, 50\}$, so that we consider cases where the used cap is above, below or equal to the true cap $k$. Unlike the standard spatial+, this allows us to control the frequencies included in the spatial effects for $\yb$ and $\xb$, such that ideally $\rb^x$ should not have any confounded frequencies. Here, on the first stage regression for $\xb$ we use no smoothing in order to guarantee the residuals $\rb^x$ are orthogonal to the subspace spanned by $\BBstilde \bbetastilde$.
We use $n$ basis functions in the spatial model.

\begin{figure}
   \centering
   \begin{minipage}{1\textwidth}
       \centering
       \includegraphics[width=1\textwidth]{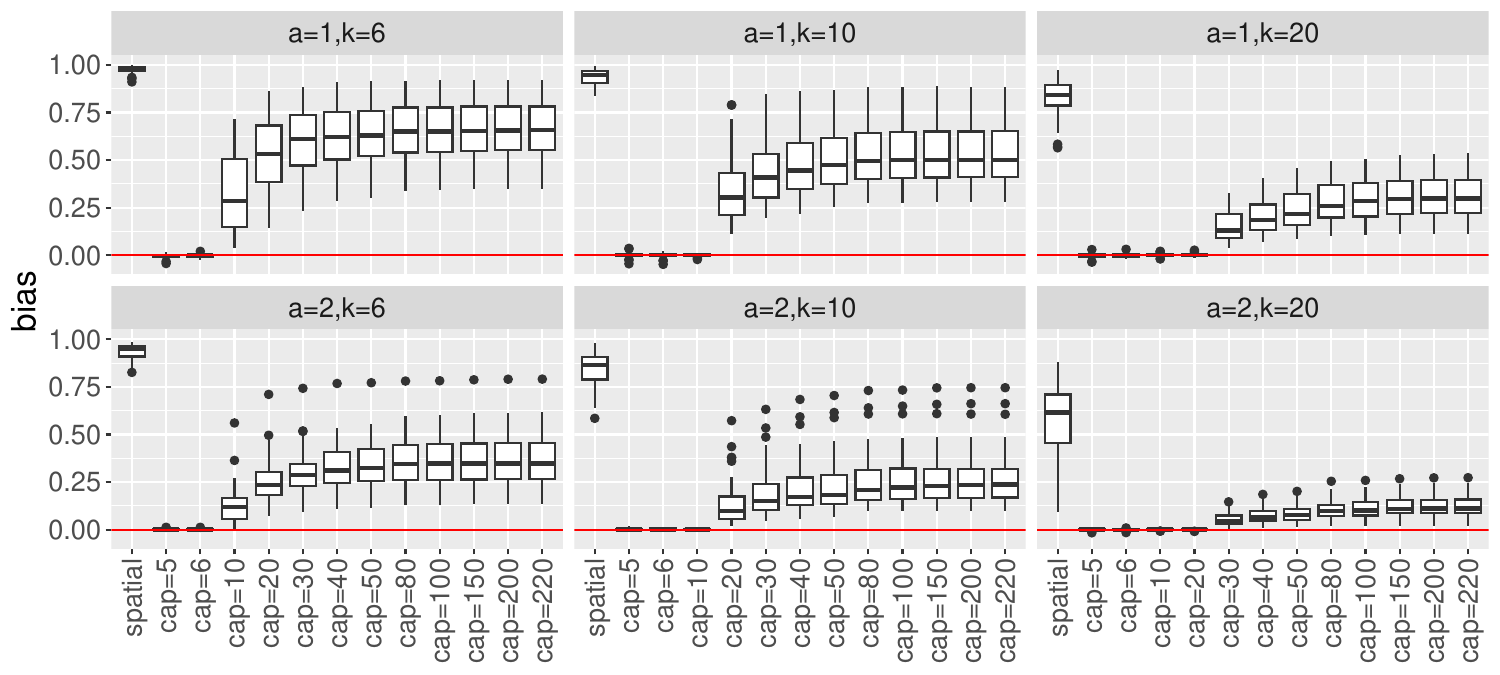} 
   \end{minipage}
   
   \caption{Bias of $\betah$ in the spatial and capped spatial+ models with different combinations of $k\in \{5,10,20\}$ and $a\in \{1, 2\}$. The true value is $\beta = 1$. All other parameters remain constant.}
   \label{fig:s6}
\end{figure}

In Figure~\ref{fig:s6} we see that the bias decreases for increasing $k$ and $a$. This is expected from our results as a larger $k$ implies fewer confounded medium frequencies and more unconfounded high frequencies, while a larger $a$ implies a larger proportion of unconfounded high frequencies in $\xb$. For $\mathrm{cap} \leq k$, we are able to completely remove confounding bias, even if only using a smaller fraction of the unconfounded high frequency component in $\xb$ to identify the effect. 
When the $k < \mathrm{cap}$ the residuals in spatial+ contain medium frequency confounding and, quite noticeably, bias arises.
Thus, we observe a constant bias of zero for caps below $k$ and a considerable increase for caps above $k$. 
The MSE for the predicted values reveals that in general capping increases MSE, although quite negligibly for $k \leq \mathrm{cap}$, when compared to the drastic increase observed for $k > \mathrm{cap}$ (see Appendix~3.1). %\ref{app:mse}).

\section{Application to air temperature in Germany}\label{sec:application}

As an empirical illustration, we analyse the relationship between the monthly mean temperature at two metres above ground, treated here as the response, and the monthly mean precipitation, the covariate, in Germany for the year 2010 at 336 locations. 
The data are open access and provided by the German Meteorological Service (DWD). 
For convenience, precipitation is measured in units of 10 millimetres and temperature is measured in degrees Celsius. 
In the context of spatial confounding, climate is a compelling example, as both covariates and response can exhibit relatively larger or narrower spatial correlation ranges, also depending on the time of year.

% n the Northern Hemisphere, including regions like Germany, the relationship between temperature and rainfall is generally weaker in winter compared to summer. This is because winter weather patterns are influenced more by large-scale weather systems, which dominate the climate, whereas summer often brings more stable weather patterns that result in more predictable and consistent rainfall. During winter, the influence of large-scale weather systems, along with topographical and localized effects, creates a more variable and irregular rainfall distribution. 

Independent regression models are fitted to each month, where rainfall and temperature are first standardised and the resulting regression coefficient for rainfall is converted back to the original scale. We consider regression models: (1) capped spatial+ (see Section~\ref{sec:capped_splus}), (2) spatial, (3) non-spatial. By comparing the results of (1) and (2), we show how capped spatial+ can help identify problematic covariate effect estimates in the spatial model and suggest appropriate adjustments.

\begin{figure} 
    \centering
    \includegraphics[width=\textwidth]{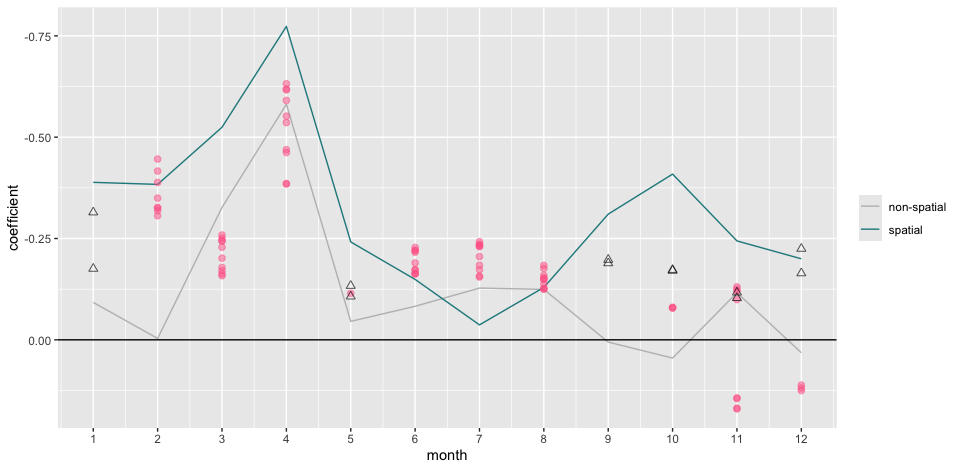}
    \caption{Coefficient estimates of $\hat{\beta}$. The solid lines correspond to the spatial model (teal), the non-spatial model (grey) and a coefficient estimate of zero (black). The pink dots show the statistically significant at 5\% estimates of the capped spatial+ for the first 15 caps. The triangles show the first two statistically significant results from capped spatial+ using the sliding window.}
    \label{fig:coefficients}
\end{figure}

All spatial effects are modelled with thin plate regression splines 
in \texttt{mgcv} \citep{mgcv}. The results are shown in Figure~\ref{fig:coefficients}. When the cap used for capped spatial+ is very small, estimates of the covariate effect may not be reliable, as the residuals will be very small and unlikely to have sufficient information about the covariate. On the other hand, as the cap increases, the fit of the model deteriorates. Therefore, we only show the estimates with $cap = 5, 6, \ldots, 15$ %there is likely little spatial information left in the residuals)
 which are also statistically significant at the 95\% confidence level. 
From September to January and May, the caps (below 15) are generally insignificant, as indicated by the lack of pink dots in the plot. 
In the other months, the capped spatial+ estimates can be used to assess whether we have confidence in the spatial estimate, and the dots show how the estimate should likely be adjusted. November is perhaps the most puzzling, where the coefficients appear to be either positive or negative depending on the frequencies included in capped spatial+. On closer inspection, this can be explained by a more complex spatial structure of precipitation in this month, with many localised patterns, so that a linear effect of precipitation may not be sufficient in general. 
With the exception of November and December, all estimates plotted in Figure~\ref{fig:coefficients} are negative, indicating a generally negative association between the two variables.

In order to understand the observed behaviour in more detail, we focus on three months: January, April and August. For these three months, results for caps up to 9, including the AIC and RMSE of the different models, are shown in Table~\ref{tab:application_results}. We start with the spring and summer months. In April, the capped spatial+ estimates are lower in absolute value than those of the spatial model. This may indicate that the spatial model estimate is biased, and under the assumption that the highest frequencies are unconfounded, the estimate should be adjusted down (in absolute terms). In August, the estimates for all models appear to be fairly consistent, with little variability in the capped spatial+ estimates. This suggests that unobserved spatial confounders are not affecting the estimation and the spatial model estimate does not need to be adjusted.
In January, we know from Figure~\ref{fig:coefficients} that none of the capped spatial+ estimates up to a cap of 15 are statistically significant. This suggests that the highest frequencies (up to 15) in the spatial effect are not helping us to identify the effect of rainfall and may simply not be present in this particular covariate. As mentioned in Section~\ref{sec:case_2}, however, when a covariate is dominated by low frequencies, the estimate in the spatial model may well be biased. In this case, we can adapt the capped spatial+ method to identify the highest frequencies at which $\xb$ has a significant effect. Assuming that these frequencies are unconfounded, we can use the corresponding capped spatial+ estimate to assess the bias. We use a sliding window technique that systematically removes 15 consecutive frequencies, starting progressively deeper into the low frequency spectrum, where $start$ is the starting frequency. Figure~\ref{fig:coefficients} shows the estimates for the first two sliding windows, with significant estimates at the 5\% level represented by triangles. We can see that in January the spatial model may overestimate the effect (in absolute terms). We followed the same approach for all the remaining autumn and winter months and for May. The results suggest that the extra peak building up to October may be an artefact, with values likely closer to flat around $-0.125$. Apart from the peak in April -- perhaps a case of ``Der April macht, was er will'' (``April does as it pleases'') -- the coefficients remain relatively stable. In all the months analysed using the sliding window technique, the first significant coefficient appears the latest at $start=8$, except in September, where it appears at $start=38$.

\begin{table}
\caption{Coefficient estimates for the regression models fitted to average monthly air temperature.}\label{tab:application_results}
\centering
\begin{tabular}{r|r|r|r|r|r}
\hline
month & model/cap & $\hat{\beta}$ & p-value & AIC & RMSE\\
\hline
1 & non-spatial & -0.0922849 & 0.0524941 & 1306.6419 & 0.9897\\
\hline
1 & spatial & -0.3885536 & 0.0000000 & 650.9834 & 0.1441\\
\hline
1 & cap =5 & -0.1847219 & 0.1679208 & 870.5588 & 0.2706\\
\hline
1 & cap =6 & 0.0654211 & 0.5980586 & 893.6362 & 0.2911\\
\hline
1 & cap =7 & 0.1690087 & 0.1617385 & 903.5738 & 0.3013\\
\hline
1 & cap =8 & 0.1773077 & 0.1480378 & 915.2859 & 0.3124\\
\hline
1 & cap =9 & 0.0240164 & 0.7846951 & 917.0572 & 0.3159\\
\hline
4 & non-spatial & -0.5811757 & 0.0000000 & 1196.6612 & 0.7748\\
\hline
4 & spatial & -0.7732586 & 0.0000000 & 951.3315 & 0.3053\\
\hline
4 & cap = 5 & -0.6189845 & 0.0000023 & 1095.9985 & 0.4349\\
\hline
4 & cap = 6 & -0.5906941 & 0.0000051 & 1096.8946 & 0.4380\\
\hline
4 & cap = 7 & -0.6318752 & 0.0000005 & 1098.1054 & 0.4404\\
\hline
4 & cap = 8 & -0.6166930 & 0.0000002 & 1096.3102 & 0.4387\\
\hline
4 &  cap = 9 & -0.3856196 & 0.0013258 & 1148.0541 & 0.5282\\
\hline
8 & non-spatial & -0.1244348 & 0.0000000 & 1231.8161 & 0.8173\\
\hline
8 & spatial & -0.1292622 & 0.0000000 & 1012.2477 & 0.3297\\
\hline
8 & cap =5 & -0.1382547 & 0.0000314 & 1157.0444 & 0.5120\\
\hline
8 & cap =6 & -0.1262546 & 0.0000555 & 1156.6078 & 0.5134\\
\hline
8 & cap =7 & -0.1248357 & 0.0000148 & 1155.2212 & 0.5126\\
\hline
8 & cap =8 & -0.1270226 & 0.0000038 & 1153.4515 & 0.5106\\
\hline
8 & cap =9 & -0.1530327 & 0.0000000 & 1154.1968 & 0.5184\\
\end{tabular}
\end{table}

\section{Discussion}
\label{s:discuss}
As illustrated in this paper, the analytical expressions derived in Proposition~\ref{prop:bias} and Corollary~\ref{cor:bias} provide an intuitive framework for explaining and, indeed, demystifying the bias arising from spatial confounding. In particular, we show that bias in the spatial model is caused by spatial smoothing and can become arbitrarily large. In practical applications, the difference between the estimates in the spatial and non-spatial models is commonly seen as an indicator of spatial confounding. However, our analysis and simulations show that a difference between these estimates does not in itself show whether the estimate in the spatial model is biased. %However, our analysis and simulations show that this difference is not directly related to the size or occurrence of bias in the spatial model. 
Instead we propose using the methods set out in Section~\ref{sec:adj} to assess and adjust for the bias.

As most other work on spatial confounding, the focus in this paper has been on quantifying the bias in spatial regressions. A derivation and short discussion about the variance of the covariate effect estimate is included in Appendix~1.6. %Appendix~\ref{app:var}
However, to put the bias into perspective and thoroughly conduct statistical inference, more investigations on associated uncertainty will be an important direction for future research.

% %  The \backmatter command formats the subsequent headings so that they
% %  are in the journal style.  Please keep this command in your document
% %  in this position, right after the final section of the main part of 
% %  the paper and right before the Acknowledgements, Supporting Information (Supplementary %  Materials),   and References sections. 

% \backmatter

% %  This section is optional.  Here is where you will want to cite
% %  grants, people who helped with the paper, etc.  But keep it short!

\section*{Data availability statement}
The data is publicly available on the website of the German weather service (DWD).  
We provide code for all analyses in the paper, which includes instructions on how to download the data.
% % \section*{Acknowledgements}

% % The authors thank Professor A. Sen for some helpful suggestions,
% % Dr C. R. Rangarajan for a critical reading of the original version of the
% % paper, and an anonymous referee for very useful comments that improved
% % the presentation of the paper.\vspace*{-8pt}

% %  Here, we create the bibliographic entries manually, following the
% %  journal style.  If you use this method or use natbib, PLEASE PAY
% %  CAREFUL ATTENTION TO THE BIBLIOGRAPHIC STYLE IN A RECENT ISSUE OF
% %  THE JOURNAL AND FOLLOW IT!  Failure to follow stylistic conventions
% %  just lengthens the time spend copyediting your paper and hence its
% %  position in the publication queue should it be accepted.

% %  We greatly prefer that you incorporate the references for your
% %  article into the body of the article as we have done here 
% %  (you can use natbib or not as you choose) than use BiBTeX,
% %  so that your article is self-contained in one file.
% %  If you do use BiBTeX, please use the .bst file that comes with 
% %  the distribution.  In this case, replace the thebibliography
% %  environment below by 
% %

% %  If your paper refers to supporting web material, then you MUST
% %  include this section!!  See Instructions for Authors at the journal
% %  website http://www.biometrics.tibs.org

% % \appendix

%  To get the journal style of heading for an appendix, mimic the following.

\label{lastpage}

\newpage
\includepdf[pages=-]{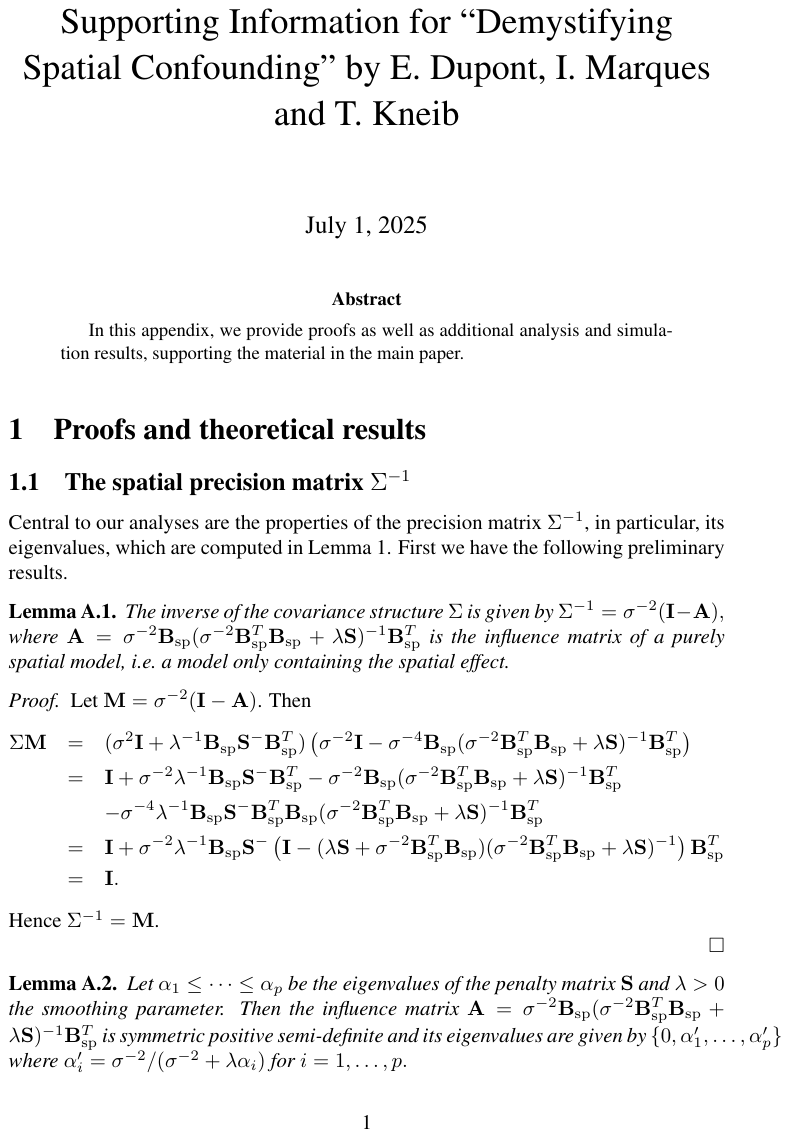}

\end{document}